# Superconducting phase diagram in Bi$_x$Ni$_{1-x}$ thin films: the effects of Bi stoichiometry on superconductivity

Jihun Park[1,2], Jarryd A. Horn[2], Dylan J. Kirsch[1,3], Rohit K. Pant[1,2], Hyeok Yoon[2], Sungha Baek[2], Suchismita Sarker[4,5], Apurva Mehta[4], Xiaohang Zhang[1], Seunghun Lee[6], Richard Greene[2], Johnpierre Paglione[2,7] and Ichiro Takeuchi[1,2,*]

[1]Department of Materials Science and Engineering, University of Maryland, College Park, MD 20742, USA
[2]Maryland Quantum Materials Center, Department of Physics, University of Maryland, College Park, MD 20742, USA
[3]Material Measurement Laboratory, National Institute of Standards and Technology, Gaithersburg, MD 20899, USA
[4]Stanford Synchrotron Radiation Lightsource, SLAC National Accelerator Laboratory, Menlo Park, CA 94025, USA
[5]Cornell High Energy Synchrotron Source, Cornell University, Ithaca, NY 14853, USA
[6]Department of Physics, Pukyong National University, Busan 48513, Republic of Korea
[7]Canadian Institute for Advanced Research, Toronto, Ontario M5G 1Z8, Canada
*Corresponding author: takeuchi@umd.edu

The Bi–Ni binary system has been of interest due to possible unconventional superconductivity aroused therein, such as time-reversal symmetry breaking in Bi/Ni bilayers or the coexistence of superconductivity and ferromagnetism in Bi$_3$Ni crystals. While Ni acts as a ferromagnetic element in such systems, the role of strong spin-orbit coupling element Bi in superconductivity has remained unexplored. In this work, we systematically studied the effects of Bi stoichiometry on the superconductivity of Bi$_x$Ni$_{1-x}$ thin films ($x \approx 0.5$ to $0.9$) fabricated via a composition-spread approach. The superconducting phase map of Bi$_x$Ni$_{1-x}$ thin films exhibited a superconducting composition region attributable to the intermetallic Bi$_3$Ni phase with different amount of excess Bi, revealed by synchrotron X-ray diffraction analysis. Interestingly, the mixed phase region with Bi$_3$Ni and Bi showed unusual increases in the superconducting transition temperature and residual resistance ratio as more Bi impurities were included, with the maximum $T_c$ (= 4.2 K) observed at $x \approx 0.79$. A correlation analysis of structural, electrical, and magneto-transport characteristics across the composition variation revealed that the unusual superconducting "dome" is due to two competing roles of Bi: impurity scattering and carrier doping. We found that the carrier doping effect is dominant in the mild doping regime ($0.74 \leq x \leq 0.79$), while impurity scattering becomes more pronounced at larger Bi stoichiometry.

## I. INTRODUCTION

Bismuth (Bi) is a heavy element characterized by strong spin-orbit coupling, with its pure phase or compounds showing a variety of superconducting phenomena, such as high-temperature, high-pressure, and topological superconductivity [1–6]. One of the exotic superconducting systems containing Bi is the Bi–Ni binary system, where alloying Bi with Ni can create two superconducting intermetallics, BiNi and Bi$_3$Ni, with $T_c \approx 4$ K for both phases [6–9]. The Bi–Ni phase diagram also encompasses a ferromagnetic Ni phase, which is typically incompatible with superconductivity due to Cooper pair breaking [10, 11]. Of particular interest in this binary system is the emergence of unconventional superconductivity, such as possible spin-triplet superconductivity with broken time-reversal symmetry in Bi/Ni bilayers [5, 12–14] and the coexistence of superconductivity and ferromagnetism in Bi$_3$Ni crystals [10, 15–19].

While it is evident that Ni acts as a magnetic element in such systems, the role of Bi in superconducting properties has yet to be fully understood. In Bi/Ni bilayer studies, superconductivity has been observed, which strongly depends on the thickness ratio of the Bi layer to the Ni layer with the $T_c$ increasing from $\approx 1.7$ K (Bi:Ni = 5) to $\approx 4.1$ K (Bi:Ni = 12) [5, 20, 21]. Also, the polar Kerr signals as a signature of time-reversal symmetry breaking have been observed, but only on the Bi side (20 nm to 40 nm) and not on the Ni side (2 nm to 6 nm) [5]. Several mechanisms have been proposed to explain this unusual superconductivity in Bi/Ni bilayers, including interfacial superconductivity [22] or proximitized superconductivity in the pure Bi layer due to Bi$_3$Ni formation [23]. Although no agreement on the mechanism has been reached so far, the pure Bi phase seems to play a vital role in the superconductivity of the Bi/Ni bilayer system.

In the case of Bi$_3$Ni, the emergence of ferromagnetism has been discussed due to magnetic ordering from stoichiometric defects, such as Ni inclusions [10]. Such disorder can be readily created during crystal growth owing to the low melting temperature and high diffusivity of elemental Bi [6, 24–26]. Interestingly, most Bi$_3$Ni materials from previous work contained trace amounts of elemental Bi as an impurity phase, as reported in various types of Bi$_3$Ni samples, including single crystals [5, 10], polycrystals [19], nanostructures [18, 26], thin films [27], Fe-doped Bi$_3$Ni [16], and Bi-deficient Bi$_3$Ni [20]. Nevertheless, the effects of the Bi inclusion are yet to be understood.

In this work, we have systematically investigated the superconducting properties of the Bi$_x$Ni$_{1-x}$ thin films using a composition-spread approach. Temperature-dependent transport characterization has allowed us to map the superconducting phase diagram of the Bi–Ni library as a continuous function of the Bi stoichiometry. An unusual enhancement in the critical temperature in Bi$_3$Ni with excess Bi inclusion has been observed. We have performed correlation analysis on structural, electronic, and magneto-transport properties of composition-spread thin films in order to uncover the role of Bi in superconductivity. Our work aims to explore superconducting behavior depending on the Bi stoichiometry, thus providing useful guidance for future studies about the Bi–Ni binary system, including Bi$_3$Ni crystals and Bi/Ni bilayers.

## II. EXPERIMENT

The composition spread libraries of Bi$_x$Ni$_{1-x}$ thin films were fabricated by co-sputtering Bi and Ni targets on 76 mm (3 inch) diameter SiO$_2$/Si substrates at room temperature, as shown in Fig. 1(a). The chamber base pressure was approximately $1 \times 10^{-5}$ Pa, and the deposition was performed under the pressure of $\approx 1.3$ Pa (10 mTorr) of Ar (99.999 %). The incident angles of two sputtering guns onto the substrate were adjusted to obtain a broad composition range of the Bi–Ni thin-film library, which was measured via wavelength dispersive spectroscopy (WDS) using an electron probe microanalyzer (EPMA; JXA 8900R Microprobe, JEOL)[1]. The composition range mapped on the Bi$_x$Ni$_{1-x}$ composition-spread libraries was found to be $x \approx 0.5$ to 0.9 (Fig. S1(a)). The uncertainty in the atomic ratio from the WDS measurements was estimated to be ± 1 % (1 sigma), primarily from the statistical analysis of the instrumentation counts. As we see below, because we have several multi-phase regions within the spread, the composition parameter $x$ here corresponds to the overall average Bi concentration at each position on the spread. The thickness of Bi$_x$Ni$_{1-x}$ thin films in the library was measured using cross-sectional scanning electron microscopy, and it was found to vary continuously from 100 nm (Bi$_{0.5}$Ni$_{0.5}$) to 200 nm (Bi$_{0.9}$Ni$_{0.1}$) resulting from different target-to-substrate distances and deposition rates between Bi and Ni in the co-sputtering process. Spectroscopic ellipsometry (SE; M-2000D, J. A. Woollam) was utilized to assess optical constants and to optically identify the phase boundaries in the Bi$_x$Ni$_{1-x}$ thin-film library. The electrical and magneto-transport properties of thin-film samples were investigated by the standard four-probe method using a Physical Property Measurement System (Quantum Design). The uncertainty in the electrical resistivity is ± 0.1 % (1 sigma) based on the instrument specifications. Structural phase distribution information was obtained using synchrotron X-ray diffraction (XRD) at Stanford Synchrotron Radiation Laboratory Beamline 1–5. A grazing incidence angle of 1 ° to 2 ° was employed to minimize the contribution of silicon substrates to the diffraction data [28]. The diffraction intensity of the composition spread was acquired as a function of the scattering vector $Q = 4\pi \sin\theta/\lambda$, where $\theta$ is the Bragg angle and $\lambda$ is the wavelength of the incident X-ray.

## III. RESULTS AND DISCUSSION

The crystal structure analysis of the library was performed to identify the distribution of phases and phase boundaries across the composition spread (Fig. 1(a)). Synchrotron XRD patterns measured at different compositions on the Bi$_x$Ni$_{1-x}$ thin-film spread are shown in a waterfall plot in Fig. 1(b) with known diffraction patterns for Bi, Bi$_3$Ni, and BiNi crystal structures displayed below for comparison. Based on their matches, three phases are identified in the library: BiNi (space group 8, $C1m1$), Bi$_3$Ni (space group 61, $Pnma$), and Bi (space group 166, $R\bar{3}m$). Further quantitative analysis based on XRD peak fitting and Rietveld refinement allowed for identification of phase boundaries in the library. The bottom of Fig. 1(a) includes a schematic view of the phase information, where Regions A (BiNi), C (Bi$_3$Ni), and E (Bi) indicate single phase regions, while B (BiNi + Bi$_3$Ni) and D (Bi$_3$Ni + Bi) correspond to mixed phase regions. The structural phase boundaries of the Bi–Ni library can be observed at $x \approx 0.67$, 0.73, 0.74, and 0.92, which are consistent with the phase boundaries identified by optical characterization using SE (Fig. S1 (b)–(d)) and by visual inspection (Fig. S2).

The presence of two mix-phase regions (Regions B and D) and three single-phase regions (A (BiNi), C (Bi$_3$Ni), and E (Bi)) are consistent with the thermodynamic bulk phase diagram, as shown in Fig. 1(a). However, the Bi concentration ($x$) corresponding to the phase boundaries differs between our thin-film spread and the bulk phase diagram [7]. Specifically, the Bi–Ni library shows a broadening in the composition range of pure phase regions (Regions A (BiNi) and E (Bi)), likely due to the non-equilibrium sputtering deposition process used for the thin-film deposition.

In the case of the bulk phase diagram, the synthesis of bulk crystals usually includes crystallization at an elevated temperature for a long time (more than a few hours) and slow cool-down procedures. These processes are governed primarily by the thermodynamics of the Bi–Ni system, such as formation energy and entropy. This thermodynamics-controlled mechanism generally leads to narrow composition ranges for intermetallic compounds (i.e., BiNi and Bi$_3$Ni), where substitutional doping is energetically more expensive than phase deconvolution with non-zero interface energy.

On the other hand, the Bi–Ni film growth using our co-sputtering approach is based on the vaporization of Bi and Ni atoms by energetic Ar plasma and subsequent rapid cooling on a Si substrate. Therefore, the sputtered atoms on the substrate do not have enough thermal energy to rearrange, thereby making our thin-film system kinetically limited (i.e., atomic diffusion for crystallization is suppressed). This kinetics-control mechanism could lead to a non-equilibrium phase diagram, including an extended range of impurity-doped single-phase regions, as shown in Regions A (BiNi) and E (Bi).

One important feature in our thin-film phase diagram is the difference between the two mixed-phase regions. That is, Region D (Bi$_3$Ni + Bi)) shows a relatively wider composition range (0.74 < $x$ < 0.92), compared to Region B (BiNi + Bi$_3$Ni; 0.67 < $x$ < 0.73). We attribute this to the difference in the diffusion coefficients between Bi ($\approx 2 \times 10^{-21}$ m$^2$s$^{-1}$) and Ni ($\approx 10^{-28}$ m$^2$s$^{-1}$) [29]. The Bi phase (Regions D or E) with mobile Bi atoms acts as a Bi reservoir, and thus Bi$_3$Ni seed crystals can be readily formed and grown even at room temperature if there are enough Ni atoms nearby. This would transform the single-phase Bi region doped with Ni atoms (Region E) into the Bi+Bi$_3$Ni mixed phase region (Region D), thus increasing the width of Region D. This is consistent with the fact that Bi$_3$Ni phase can be spontaneously created at the Bi/Ni interface as reported in previous Bi/Ni bilayer studies [6]. On the other hand, both BiNi and Bi$_3$Ni phases do not allow phase evolution of constituent atoms due to immobilized Ni atoms (i.e., low thermal diffusivity) and the large activation energy for Bi out-diffusion from the intermetallic phases. This would make the width of Region B remain unexpanded. As a result, the large diffusion constant of Bi atoms makes Region D less limited by the kinetic mechanism, thus creating the Bi$_3$Ni+Bi mixed phase region (Region D) closer to the bulk phase diagram than Region B (BiNi+Bi$_3$Ni).

---

[1] Certain commercial equipment, instruments, or materials are identified in this document. Such identification does not imply recommendation or endorsement by the National Institute of Standards and Technology, nor does it imply that the products identified are necessarily the best available for the purpose.

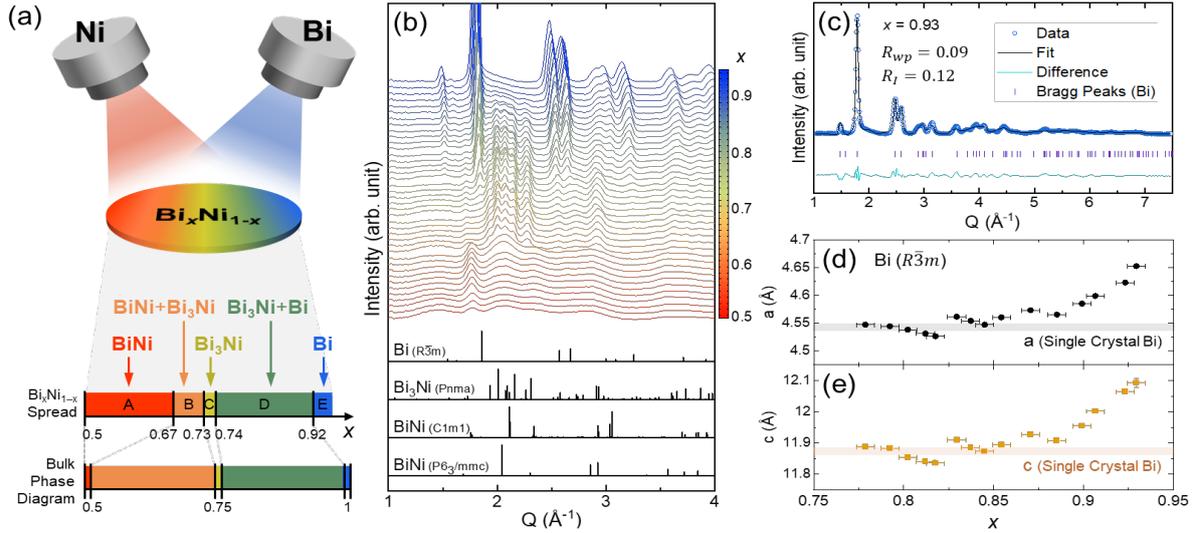

FIG 1. (a) Schematic view of the combinatorial deposition of the $Bi_xNi_{1-x}$ thin-film library with identified structural phase regions and boundaries. A bulk phase diagram is shown below for comparison. (b) Synchrotron X-ray diffraction patterns for the $Bi_xNi_{1-x}$ library. Simulated diffraction patterns for Bi ($R\bar{3}m$), $Bi_3Ni$ ($Pnma$), BiNi ($C1m1$), and BiNi ($P6_3/mmc$) are shown for phase identification below. (c) Rietveld refinement for the X-ray diffraction patterns of $Bi_{0.93}Ni_{0.07}$ thin film. The Bragg peaks correspond to the Bi ($R\bar{3}m$) diffraction patterns. $R_{wp}$ and $R_I$ refer to $R$-factors based on weighted profile and pattern intensity, respectively. Lattice parameters, $a$ and $c$, of the $Bi_xNi_{1-x}$ thin films in the library are presented in (d) and (e). The colored bars correspond to the lattice parameters of the Bi single crystal reference (JCPDS No. 44-1246).

Intriguingly, the pure $Bi_3Ni$ phase can be found at $x \approx 0.73$ to 0.74 (Region C) with a narrow composition range ($\Delta x \approx 0.01$), which appears to be slightly shifted from the expected stoichiometry composition with $x = 0.75$ rather than the region being broadened due to a doping effect. Since the shift is close to the resolution limit of our WDS measurements, it could be due to a systematic measurement uncertainty. We note that a similar experimental result was found in a growth kinetics study of Bi–Ni intermetallics, where a pure $Bi_3Ni$ phase was made in a slightly Bi-deficient region with $x \approx 0.733$ [25]. Moreover, Bi inclusion was observed even in Bi-deficient $Bi_3Ni$ crystals [20]. If meaningful, the shift can be explained by the large difference in diffusivity between Bi and Ni atoms [25, 26], leading to selective Bi out-diffusion and local precipitation of the Bi phase during the growth of $Bi_3Ni$ crystals at room temperature. In other words, only Bi can diffuse out from $Bi_3Ni$ at $x = 0.75$, and the out-diffused Bi atoms likely form Bi nanocrystals locally, leaving the $Bi_3Ni$ crystals partially deficient with Bi.

The structural properties of such Bi off-stoichiometry in $Bi_xNi_{1-x}$ thin films were further analyzed in detail. We refined the XRD patterns of the Bi-rich region on the spread ($x > 0.92$) and confirmed that Region E is purely of the Bi phase. Figure 1(c) shows the Rietveld refinement result for the $Bi_{0.93}Ni_{0.07}$ composition spot on the thin-film spread. The XRD pattern of this sample was refined well with a single Bi phase ($R\bar{3}m$). The $R$-factors based on the weighted profile ($R_{wp}$) and pattern intensity ($R_I$) were 0.09 and 0.12, respectively. Thus, no phases other than the pure Bi phase can be detected in this sample within the resolution limit of synchrotron XRD. Figures 1(d) and 1(e) present lattice parameters ($a$ and $c$) of the Bi phase found in $Bi_xNi_{1-x}$ spread thin films. While the lattice parameters of Bi inclusion in $Bi_xNi_{1-x}$ ($x < 0.9$) are comparable to those of Bi single crystals, significant lattice expansion can be seen when $x > 0.9$. The increase in the lattice parameters of the Bi phase for $x > 0.9$ is accompanied by the suppression of $Bi_3Ni$ formation, implying Ni atoms prefer to be consumed via doping into Bi crystals rather than creating the $Bi_3Ni$ phase. This agrees with a previous report where up to $\approx 10$ % of Ni can be doped in a rhombohedral Bi matrix as a solid solution [30]. Figure 2 shows the molar fractions of $Bi_3Ni$ and Bi phases in the $Bi_xNi_{1-x}$ thin films as a function of $x$. These phase fractions were obtained from quantitative analysis of synchrotron X-ray diffraction data. The thermal-equilibrium phase fractions estimated from the bulk phase diagram via the lever rule are also included (dashed lines). The Bi phase fraction in the bulk phase diagram for an extended range of $x$ is provided in Fig. S3.

Figure 3(a) shows the temperature-dependent electrical resistance ($R$–$T$) of the $Bi_xNi_{1-x}$ thin-film library with the inset displaying a zoom-in view for the $R$–$T$ curves at low temperatures ($T < 5$ K). The electrical resistance of the samples was normalized to the room-temperature resistance ($R_{300K}$) to exclude geometric effects, such as the thickness and dimensions of the films. The $R$–$T$ curve of a pure Bi thin film made under the same condition is included as a reference. The superconducting phase diagram was obtained from the temperature-dependent normalized resistance of $Bi_xNi_{1-x}$ thin films, as presented in Fig. 3(b). The phase diagram shows the superconducting phase (SC, $R = 0$) and non-superconducting phase (Non-SC, $R > 0$) regions. The phase boundary information acquired from the structural analysis is also included.

The $Bi_xNi_{1-x}$ library exhibits superconductivity with $T_c > 1.8$ K when the Bi compositions range from $x \approx 0.69$ to $x \approx 0.92$. This is reasonably consistent with Regions (B, C, and D) where the $Bi_3Ni$ phase is observed ($0.67 < x < 0.92$). Also, no sign of superconductivity was observed down to 1.8 K in single-phase Bi samples ($x > 0.92$; Region E) and a separately prepared (as a reference) pure Bi thin film. These results can rule out spurious origins of superconductivity other than that from the $Bi_3Ni$ phase, including amorphous Bi [31]. In particular, when an amorphous Bi sample is held at $T > 30$ K, its superconductivity disappears permanently due to crystallization resulting from the high diffusivity of Bi atoms [6], whereas the Bi–Ni library in this work was fabricated at room temperature. Thus, we attribute superconductivity in the Bi$_x$Ni$_{1-x}$ thin-film library solely to the superconducting Bi$_3$Ni phase.

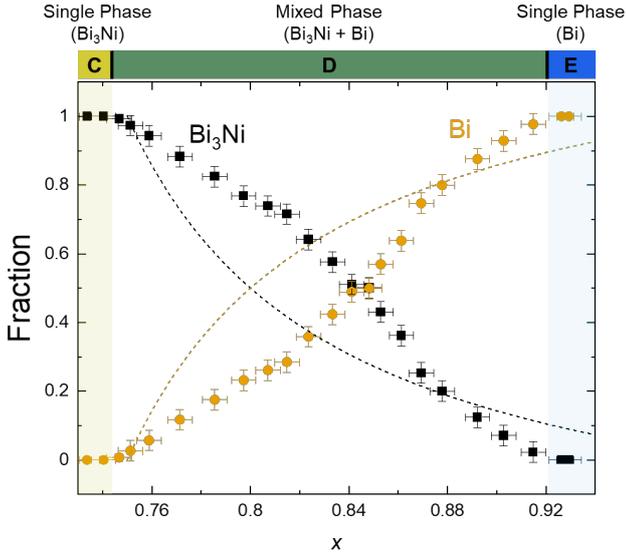

FIG 2. Molar fraction of Bi$_3$Ni and Bi phases in Bi$_x$Ni$_{1-x}$ thin films as a function of Bi compositions. Dashed lines correspond to the molar fraction of each phase calculated from the bulk phase diagram using the lever rule.

No superconducting transition was observed in the BiNi phase region (Region A). It turned out that the BiNi phase in our Bi–Ni spread library has a monoclinic structure with space group $C1m1$, while the known superconducting BiNi phase reported previously has a hexagonal symmetry with space group $P6_3/mmc$ [9]. The formation energies of the hexagonal ($E \approx -0.052$ eV/atom) and monoclinic ($E \approx -0.043$ eV/atom) phases are comparable with $\Delta E < 0.01$ eV/atom [32], indicating both are thermodynamically stable structures. Considering that the low-symmetry monoclinic structure is relatively more stable at lower pressures and higher temperatures in similar binary systems with monoclinic-hexagonal phase transitions [33–35], the energetic sputtering deposition and the subsequent rapid cooling of the films under a vacuum environment likely formed the monoclinic phase in our spread library.

Interestingly, two mixed-phase regions (B and D) show different superconducting behaviors. In Region B (BiNi + Bi$_3$Ni), superconductivity is suppressed as more amount of the impurity phase (BiNi) is included in the system, with $T_c$ gradually decreasing from 3.4 K ($x \approx 0.74$) to 2.0 K ($x \approx 0.70$). In Region D (Bi$_3$Ni + Bi), on the other hand, regardless of the fraction of Bi inclusion in Bi$_3$Ni, $T_c$ remains relatively high ($\geq 3.4$ K) with a maximum $T_c$ of 4.2 K at $x \approx 0.79$ to 0.80. In particular, the Bi$_{0.91}$Ni$_{0.09}$ sample with only $\approx 2\%$ molar fraction of Bi$_3$Ni included shows comparably robust superconductivity ($T_c \approx 3.4$ K), while superconductivity suddenly disappears at $x > 0.92$ where no Bi$_3$Ni is observed (Fig. 3). To better understand the unusual superconducting behavior in Bi$_3$Ni with Bi inclusion (Region D), we performed field-dependent transport measurements on the Bi$_x$Ni$_{1-x}$ thin films.

Figure 3(c) shows the normalized $R$–$T$ curves of the Bi$_{0.80}$Ni$_{0.20}$ thin film with different magnetic fields applied in the out-of-plane direction, where a continuous decrease in $T_c$ is observed. Fig. 3(d) presents temperature-dependent upper critical fields, $\mu_0H_{c2}(T)$, obtained from the field-dependent $R$–$T$ measurements for Bi$_x$Ni$_{1-x}$ thin films, where $\mu_0$ is the vacuum magnetic permeability. The solid and dashed lines correspond to the fitting curves obtained using the Werthamer–Helfand–Hohenberg (WHH) formula in the dirty limit [36, 37]:

$$\mu_0 H_{c2}^{\text{WHH}}(0) = -0.693 T_c \left(\frac{dH_{c2}}{dT}\right)_{T_c} \quad (1)$$

We employed an approach proposed by T. Baumgartner et al. [38] for temperature-dependent fitting of the WHH formula (1.8 K $\leq T \leq T_c$). The WHH model fitted the data well with $R^2 > 0.99$ for all superconducting Bi$_x$Ni$_{1-x}$ thin-film samples here, which was confirmed to provide better fitting results than the temperature-dependent Ginzburg–Landau model. From the WHH fit, we obtained the temperature-dependent upper critical fields, $\mu_0H_{c2}(T)$, and calculated the Ginzburg–Landau coherence length ($\xi_{GL}$) using the following equation [39]:

$$\xi_{GL}(T) = \sqrt{\frac{\Phi_0}{2\pi\mu_0 H_{c2}(T)}} \quad (2)$$

where $\Phi_0 = h/2e \approx 2.07 \times 10^{-15}$ T·m$^2$ is the magnetic flux quantum. The calculated upper critical fields and Ginzburg–Landau coherence lengths for Bi$_x$Ni$_{1-x}$ thin films are provided in Fig. 4(a). The normal-state magnetoresistance (MR) of Bi$_x$Ni$_{1-x}$ thin films, defined as $MR = [R(H) - R(0)]/R(0)$, was measured at 4.5 K (Fig. 4(b)). No MR is observed for $x \leq 0.74$ in Region B (BiNi + Bi$_3$Ni mixed phase) and Region C (Bi$_3$Ni single phase), whereas samples with $x > 0.74$ (Region D, Bi$_3$Ni + Bi) show positive MR whose magnitude is proportional to the amount of Bi inclusion. This positive MR is consistent with that of single-phase Bi crystals due to field-dependent scattering [40, 41], indicating that the Bi inclusion acts as an impurity phase in the magneto-transport of Bi$_x$Ni$_{1-x}$ thin films.

The residual resistance ratio (RRR) above the superconducting transition temperature changes with varying the Bi concentration in Regions B, C, and D. We extracted the RRR values of Bi$_x$Ni$_{1-x}$ thin films at 5 K from the $R$–$T$ measurement data, where RRR = $R_{300\text{K}}/R_{5\text{K}}$. The RRR of the spread library shows a systematic change with a maximum RRR of 1.82 at $x \approx 0.79$.

Figure 5 shows a comparison of $T_c$, RRR, and MR of Bi$_x$Ni$_{1-x}$ thin films as a function of $x$. As mentioned earlier, single-phase Bi$_3$Ni thin films (Region C) show $T_c$ of $\approx 3.4$ K, which is drastically reduced as more BiNi impurities are included (Region B), as marked with arrow "a" in Fig. 5(a). Meanwhile, samples in Region D (Bi$_3$Ni + Bi mixed phase) show relatively high $T_c$ for all compositions, exhibiting a first increasing (arrows "b") and then decreasing (arrow "c") trend with the maximum $T_c$ at $x \approx 0.79$. We note that the changing trend of $T_c$ according to $x$ is similar to that of RRR, shown in Fig. 5(b), indicating that $T_c$ and RRR are correlated. Figure 5(c) shows the MR of Bi$_x$Ni$_{1-x}$ thin films at $H = 9$T, defined as $MR_{9T} = [R(9T) - R(0T)]/R(0T)$. The normal-state MR drastically increases for samples with $x > 0.79$, where a decreasing trend is observed for both $T_c$ and RRR. This suggests that the reduction in RRR and $T_c$ (arrow "c") is due to the significant amount of Bi inclusion acting as an impurity phase in the films. We note that the thickness variation in our spread would not affect these trends significantly, as RRR and MR are normalized values where the dimensional effect is removed. Also, the thickness variation is only ± 10 % (180 ± 20 nm) in the composition range where the most dramatic changes in the trends are observed ($0.7 \leq x \leq 0.85$).

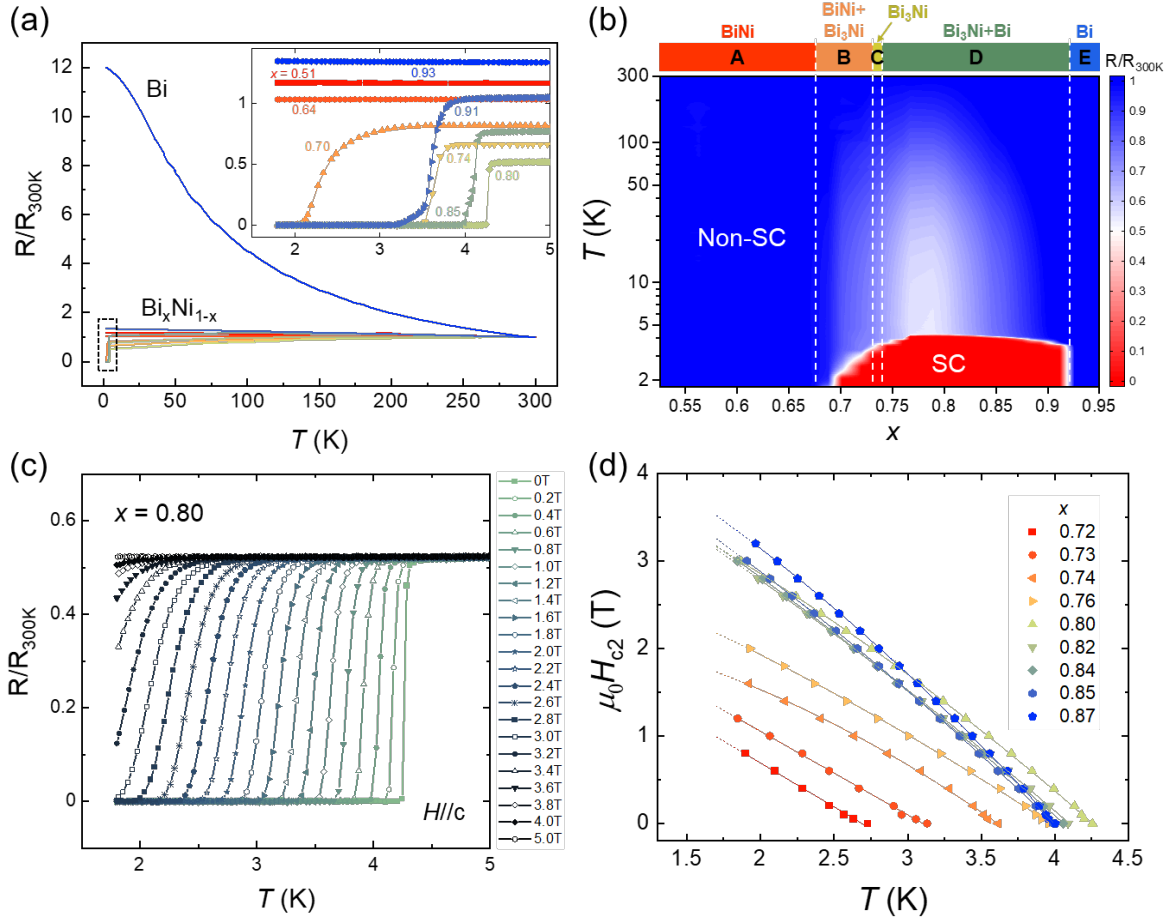

FIG 3. (a) Normalized resistance of pure Bi and $Bi_xNi_{1-x}$ thin films. Inset shows low-temperature data for the $Bi_xNi_{1-x}$ library, corresponding to the marked dotted region. (b) Superconducting phase diagram of $Bi_xNi_{1-x}$ thin films, obtained from the normalized resistance as functions of temperature and the total Bi composition ($x$). The phase boundary information (A to E) is included for reference. The red area indicates zero-resistance region, corresponding to superconducting phase (SC), while blue and white areas show non-superconducting regions (non-SC) with $R/R_{300K} > 0$. (c) Field-dependent $R/R_{300K}(T)$ plots for $Bi_{0.8}Ni_{0.2}$ thin film. (d) The upper critical fields of $Bi_xNi_{1-x}$ thin films as a function of temperature. Solid and dashed lines correspond to the fitted curves based on the Werthamer–Helfand–Hohenberg (WHH) equation for interpolation and extrapolation, respectively.

Here, the question needs to be addressed as to why $T_c$ and RRR increase for Bi compositions corresponding to $0.74 \leq x \leq 0.79$ (arrow "b"). One possible scenario is that the Bi inclusion allows for the regrowth of $Bi_3Ni$ crystals at room temperature via the inter-diffusion of highly diffusive Bi atoms between Bi and $Bi_3Ni$ [6, 25, 26, 29]. This recrystallization due to Bi inclusion would enhance the crystal size and crystallinity of the $Bi_3Ni$ phase in the films, thus decreasing grain boundary scattering and increasing carrier mobility and the mean free path. We performed further correlation analysis to verify this scenario and elucidate the abnormal superconductivity increase in the arrow "b" region.

Figure 6(a) shows the crystallite size ($D$) and carrier mean free path ($l_{MFP}$) of $Bi_3Ni$ crystals in $Bi_xNi_{1-x}$ thin films. The crystallite size was calculated using the Scherrer equation [42], $D = K\lambda/\beta \cos\theta$, where $K = 0.9$ is a shape factor, $\lambda$ is the wavelength of the synchrotron X-ray beam, $\beta$ is the full width at half-maximum of the diffraction peaks, and $\theta$ is the Bragg angle. The carrier mean free path $l_{MFP}$ was obtained from the following Ginzburg-Landau relation [39, 43],

$$\xi_{GL} = 0.739(\xi_0^{-2} + 0.882(\xi_0 l_{MFP})^{-1})^{-1/2}(1 - T/T_c)^{-1/2} \quad (3)$$

where $\xi_0 = \hbar v_F/\pi\Delta \approx 21$ nm is the Bardeen-Cooper-Schrieffer (BCS) coherence length of $Bi_3Ni$ crystals estimated using the Fermi velocity ($v_F$) and the superconducting energy gap ($\Delta$) values from previous reports [44, 45].

As shown in Fig. 6(a), the crystallite size of the single-phase $Bi_3Ni$ ($x \approx 0.73$) is $D \approx 17$ nm, which does not vary significantly as the Bi composition increases. In contrast, the $l_{MFP}$ of $Bi_3Ni$ gradually decreases from 17 nm ($x \approx 0.73$) to 5 nm ($x \approx 0.89$) as more Bi is included. Thus, both $D$ and $l_{MFP}$ as a function of $x$ is not consistent with the idea that the increase in RRR and $T_c$ (arrow "b") can be attributed to the film crystallinity enhancement.

Notably, $l_{MFP}$ at $x \approx 0.73$ (Region C, $Bi_3Ni$ single phase) is comparable to the size of $Bi_3Ni$ crystallites, indicative of grain boundary scattering as a dominant carrier scattering mechanism for the single-phase $Bi_3Ni$. However, $l_{MFP}$ decreases drastically as $x$ increases when $0.74 \leq x \leq 0.79$. This suggests that scattering due to impurity doping comes into play as an additional scattering mechanism. We calculated the mean free path values using the Matthiessen's rule [43], $l_{MFP}^{-1} = l_{Bi_3Ni}^{-1} + l_{imp}^{-1}$, where $l_{Bi_3Ni}$ is the mean free path due to the grain boundary scattering in Bi$_3$Ni crystals and $l_{imp}$ is the mean free path due to impurity scattering. We obtained $l_{Bi_3Ni} \approx D \approx 17$ nm for the single-phase Bi$_3$Ni ($x \approx 0.73$) indicative of no impurity scattering, while the total carrier mean free path is changed significantly due to Bi doping for $x \approx 0.79$ ($l_{MFP} \approx 7$ nm and $l_{imp} \approx 12$ nm).

To prove whether Bi atoms are indeed doped in Bi$_3$Ni, we analyzed the inter-planar spacing of Bi$_3$Ni crystals. The inter-planar spacing of (304) planes in Bi$_3$Ni crystals, $d_{(304)}$, was calculated using the (304) diffraction peaks because they have no overlap with other peaks, thus enabling accurate extraction of peak positions. $d_{(304)}$ was found to increase significantly from $2.888 \pm 0.001$ Å ($x \approx 0.73$) to $2.898 \pm 0.001$ Å ($x \approx 0.79$). Since the atomic radius of Bi (160 pm) is larger than that of Ni (135 pm) [46], the increased $d_{(304)}$ indicates the doping of Bi atoms into Bi$_3$Ni, in good agreement with the impurity doping effects on the mean free path as discussed above. These results suggest that the Bi impurity doping in Bi$_3$Ni could be the origin of the enhanced $T_c$ and RRR of Bi$_x$Ni$_{1-x}$ thin films observed for $0.74 \leq x \leq 0.79$ (in the arrow "b" region).

In general, impurity doping in metallic superconductors reduces RRR and $T_c$ by increasing carrier scattering [47–49]. On the contrary, doping impurities into superconductors with poor metallicity can enhance RRR and/or $T_c$ by increasing carrier concentrations via the change in the electronic structures, such as energy band filling and overlapping [50–54]. Such enhanced superconductivity due to carrier doping is often observed in strongly correlated systems, such as the high-$T_c$ cuprates and iron-based superconductors [55–57]. In particular, an enhancement in RRR and superconductivity due to carrier doping can be accompanied by metal-insulator-like transitions in correlated systems with low conductivity, where the RRR values are approximately unity [58].

To explore the possibility of carrier doping, we estimated the normal-state carrier concentration ($n$) of the Bi$_3$Ni phase in our Bi–Ni spread using the transport equation, $n = m^*\sigma/e^2\tau = m^*v_F\sigma/e^2l_{MFP}$, where $\sigma$ is the electrical conductivity, $m^* = 7.3 \times 10^{-30}$ kg is the effective mass of charge carriers in the Bi$_3$Ni crystals, and $\tau$ is the relaxation time depending on the mean free path [39, 44, 45]. For this estimation, we assumed the electrical current flows predominantly through the Bi$_3$Ni crystals in Bi$_x$Ni$_{1-x}$ thin films since the conductivity of single-phase Bi$_3$Ni ($\sigma_{Bi_3Ni} = 4.6 \times 10^3$ S·cm$^{-1}$; Region C) is larger by three orders of magnitude than that of pure Bi ($\sigma_{Bi} = 5.1 \times 10^0$ S·cm$^{-1}$). The positive MRs of Bi$_x$Ni$_{1-x}$ thin films induced by the Bi phase, which are smaller by two to three orders of magnitude than pure Bi crystals [39, 40], can be supportive of this assumption. The normal-state carrier concentrations of Bi$_x$Ni$_{1-x}$ thin films estimated at $T \approx 5$ K change systematically according to $x$, with a maximum $n \approx 2.3 \times 10^{21}$ cm$^{-3}$ at $x \approx 0.79$, as shown in Fig. 6(b). As more Bi is included from $x \approx 0.79$, $n$ gradually decreases down to $\approx 3.6 \times 10^{20}$ cm$^{-3}$. This decrease for $x > 0.79$ is likely due to the reduced volume fraction of metallic Bi$_3$Ni with more Bi phase added, thereby diluting carrier concentrations in the Bi$_x$Ni$_{1-x}$ thin films. We also estimated the carrier concentration of the pure Bi thin film at 5 K ($n \approx 3.6 \times 10^{17}$ cm$^{-3}$), which is three to four orders of magnitude lower than that of Bi$_3$Ni. The estimated carrier concentration of Bi is consistent with the previous reports ($n \approx 3 \times 10^{17}$ cm$^{-3}$ at 4.2 K) [59, 60]. Thus, the carrier doping scenario agrees well with the change in RRR and $T_c$ (arrow "b") in Figs. 5(a) and (b).

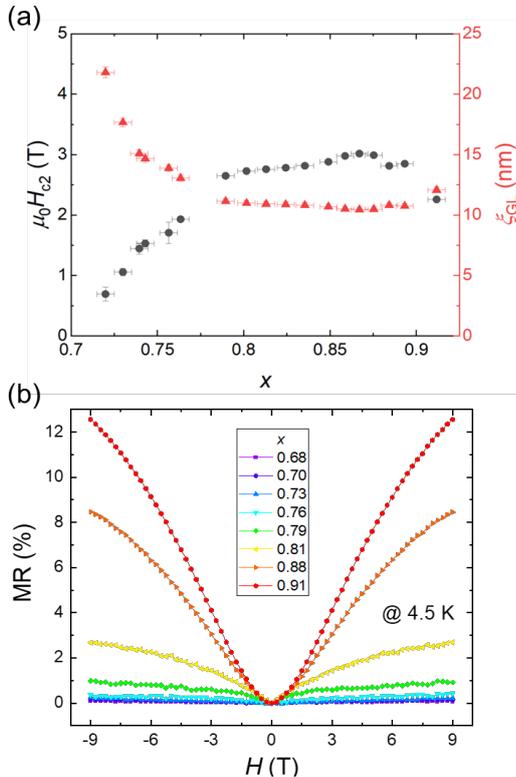

FIG. 4. (a) Upper critical field and Ginzburg-Landau coherence length of Bi$_x$Ni$_{1-x}$ thin films estimated at 2 K. (b) Magnetoresistance of normal-state Bi$_x$Ni$_{1-x}$ thin films measured at 4.5 K.

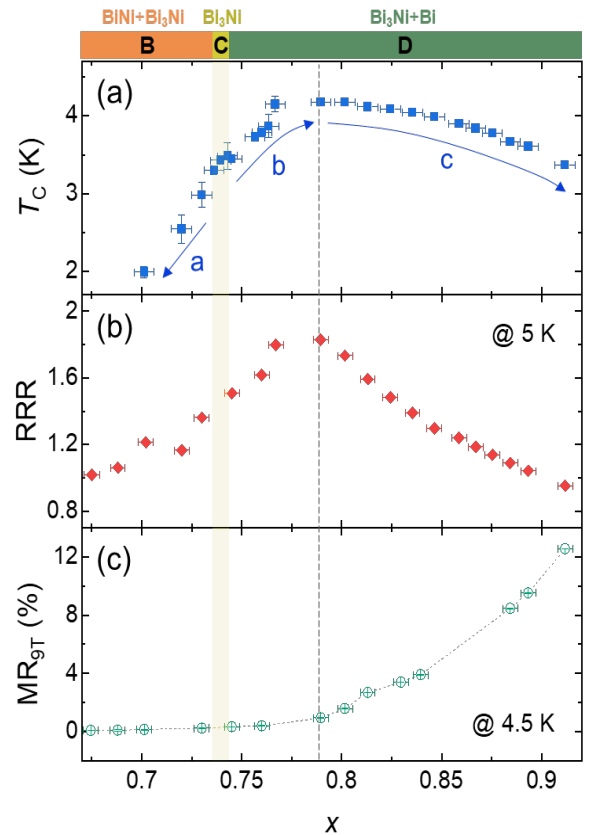

FIG 5. (a) Superconducting critical temperatures ($T_c$), (b) residual resistance ratios (RRR), and (c) magnetoresistance at 9 T (MR$_{9T}$) of Bi$_x$Ni$_{1-x}$ thin films. Phase information (Regions B to D) identified from the structural analysis is included. Error bars correspond to the standard deviations of measurement data.

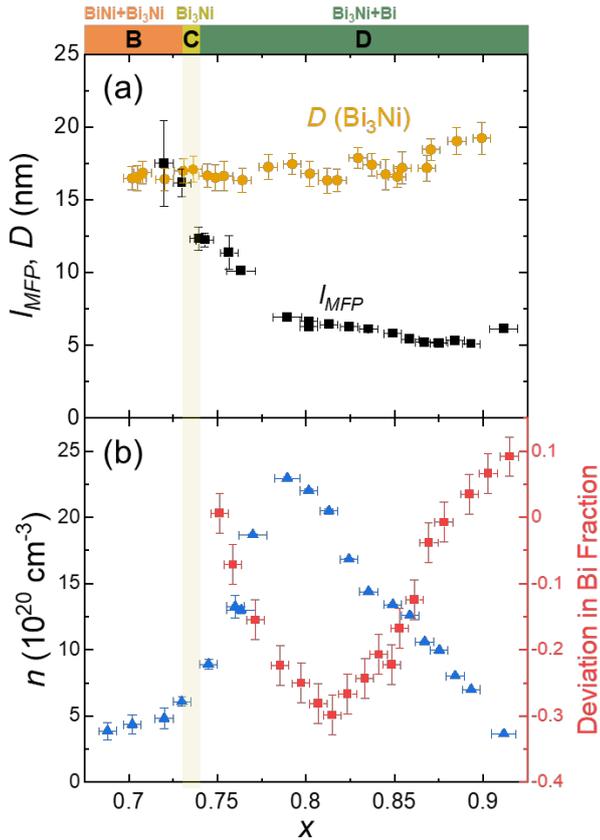

FIG. 6. (a) Crystallite size ($D$) and mean free path ($l_{MFP}$) of the superconducting $Bi_3Ni$ phase in the $Bi_xNi_{1-x}$ thin-film library extracted from synchrotron diffraction patterns. (b) Normal-state carrier concentration of $Bi_xNi_{1-x}$ thin films at 5 K. The difference between the bulk and thin-film Bi phase fractions is provided in red. Error bars were estimated from the standard deviations of measurement quantities for calculations.

We estimated the amount of Bi impurity doping in the $Bi_3Ni$ phase using the phase fractions obtained via the quantitative analysis results shown in Fig. 2. The orange dashed line in Fig. 2 shows the fraction of the Bi phase in the case of the bulk phase diagram, which corresponds to the thermal equilibrium state. The fractions of the Bi phase in our thin-film samples correspond to non-equilibrium states and thus show deviations from the bulk phase curve. The difference is mainly due to the impurity doping effect, such as Bi impurity doping into $Bi_3Ni$ phase. We estimated the difference between the bulk and thin-film Bi phase fractions, defined as $\Delta\chi_{Bi} = \chi_{Bi(thin\ film)} - \chi_{Bi(bulk)}$. Here, the signs of $\Delta\chi_{Bi}$ indicate possible types of impurity doping, where $\Delta\chi_{Bi} < 0$ and $\Delta\chi_{Bi} > 0$ cases correspond to an insufficient fraction of the Bi phase due to Bi doping in $Bi_3Ni$ and loss of $Bi_3Ni$ phase fraction due to the formation of the Ni-doped Bi phase, respectively. The thin-film phase fractions are identical to those of the thermal equilibrium case when $\Delta\chi_{Bi} = 0$. Surprisingly, the trend of $\Delta\chi_{Bi}$ is exactly opposite to that of the carrier concentration with maximum Bi impurity doping observed at $x \approx 0.81$, as shown in Figure 6(b). Also, the $\Delta\chi_{Bi}$ plot allows for the observation of the sign change at $x \approx 0.88$, reasonably consistent with the Ni-doped Bi phase identified with the lattice parameter analysis in Figs. 1(d) and (e). These results support our analysis of the Bi impurity doping well.

According to the Anderson's theorem, superconductivity is robust against scattering with non-magnetic impurities [49, 61], which successfully explained previous experiments on elemental metallic superconductors with high carrier concentrations ($n \sim 10^{23}$ cm$^{-3}$). However, this theorem is based on a critical assumption that the superconducting order parameter ($\Delta$), related to the density of Cooper-pair electrons, is uniform over the materials [49]. Thus, the Anderson theorem could not be applicable to systems where carrier concentrations change significantly due to impurity doping or band structure modifications. Furthermore, carrier scattering with strong spin-orbit coupling impurities, such as Bi, likely induces non-trivial effects on the Cooper pair wave function, making the system unexplainable with conventional superconductivity theories [49]. The numbers of valence electrons in Ni (+2) and Bi (+3 or +5) are different, which suggest substituting Ni with Bi in $Bi_3Ni$ would provide additional carriers per unit cell. This, in turn, likely changes the filling and overlapping of electronic band structure since $Bi_3Ni$ lies in the bad metallic regime with $n \sim 10^{20}$ cm$^{-3}$ and RRR of approximately 1. Such characteristics make the $T_c$ and RRR of $Bi_xNi_{1-x}$ thin films in the mild doping level ($0.74 \leq x \leq 0.79$) more sensitive to carrier doping effects than impurity scattering. On the other hand, $Bi_xNi_{1-x}$ thin films with excessive Bi inclusions ($x > 0.79$) shows reductions in $n$, RRR, and $T_c$, which closely tracks the significant increase in MR. These results suggest that the role of Bi as impurity phase becomes more dominant in the Bi-rich phase ($x > 0.79$).

Intriguingly, Bi impurity doping has been reported to increase $T_c$ in various other superconductors. These include Pb [43], $AgSnSe_2$ [62], $Pb_2Pd$ [63], and $BaPbO_3$ [64–66]. Many of these materials show poor metallic behaviors with RRRs close to 1 and relatively low carrier concentrations. Considering the substitution of Bi (+3 or +5) for different valence atoms, such as Sn (+2 or +4) or Pb (+2 or +4), the enhanced superconductivity in these materials is likely related to carrier doping effects, as in the Bi–Ni binary system. Due to the multivalency of host and dopant elements in such systems, including Bi, a valence-skip approach may also be relevant to the change in superconducting properties [62, 65].

In previous Bi/Ni bilayer studies, whether the superconducting mechanism is intrinsic or if it is due to an impurity phase was debated. In our work, a trace amount ($\approx 2\ \%$) of $Bi_3Ni$ nanocrystals ($D < 20$ nm) embedded in the Bi film was identified to induce bulk superconductivity ($T_c \approx 3.5$ K) in the Bi–Ni system. Furthermore, the trend of $T_c$ as a function of the Bi:Ni ratio ($= x/(1-x)$) resembles that observed in previous Bi/Ni bilayer studies (Fig. S4)[5, 20]. Thus, to completely exclude possible superconductivity origin induced by the impurity phase, future Bi/Ni bilayer studies should ensure the absence of nanoscale $Bi_3Ni$ at the interface. One possible way to do this is to fabricate and measure Bi/Ni bilayer samples under in-situ low-temperature environments where no $Bi_3Ni$ phase can form spontaneously, as reported in [6].

In addition, our results provide a reasonable explanation regarding possible Bi inclusions in $Bi_3Ni$ crystals in previous studies. We found that the superconductivity of $Bi_3Ni$ is extremely robust against the Bi impurities regardless of the impurity form (substitutional doping or impurity phase precipitation). Besides, the stoichiometric $Bi_3Ni$ ($x = 0.75$) region was found to possibly include the Bi impurity phase, which is likely due to the difference between Bi and Ni diffusivity and the resulting local precipitation. These findings provide a useful database and insight into the Bi–Ni material system for future superconductivity studies.

## IV. CONCLUSION

We performed a systematic investigation of the superconducting properties of $Bi_xNi_{1-x}$ thin films via a combinatorial thin-film library approach. Synchrotron XRD analysis revealed that the thin-film library contains BiNi, $Bi_3Ni$, and Bi phases with different phase fractions as a function of the total Bi composition across the library. An unusual superconducting "dome" was observed in the superconducting phase diagram obtained from the $R$–$T$ measurements, where the $T_c$ of $Bi_3Ni$ increases as more Bi impurities are added. This mixed phase region showed robust superconductivity even in excessively Bi-rich regions ($x \approx 0.90$), corresponding to a few percent of $Bi_3Ni$. Correlation analysis of structural, electrical, and magneto-transport characteristics suggested that two competing mechanisms (i.e., carrier doping and impurity scattering) exist in the mixed phase region of Bi and $Bi_3Ni$ with an optimal composition of $x \approx 0.79$ exhibiting the highest $T_c$, RRR, and $n$. These results provide insights into unusual superconductivity in compounds containing the element Bi.

## ACKNOWLEDGEMENTS

This work was supported by AFOSR FA9550-14-10332 and NIST grant #60NANB19D027 with partial support from Laboratory of Physical Sciences. D.J.K. acknowledges the partial financial support of the NSF Graduate Research Fellowship and the UMD Clark Doctoral Scholars Fellowship.

[ Supplementary Information ]

# Superconducting phase diagram in $Bi_xNi_{1-x}$ thin films: the effects of Bi stoichiometry on superconductivity

Jihun Park[1,2], Jarryd A. Horn[2], Dylan J. Kirsch[1,3], Rohit K. Pant[1,2], Hyeok Yoon[2], Sungha Baek[2], Suchismita Sarker[4,5], Apurva Mehta[4], Xiaohang Zhang[1], Seunghun Lee[6], Richard Greene[2], Johnpierre Paglione[2,7] and Ichiro Takeuchi[1,2,*]

[1]Department of Materials Science and Engineering, University of Maryland, College Park, MD 20742, USA

[2]Maryland Quantum Materials Center, Department of Physics, University of Maryland, College Park, MD 20742, USA

[3]Material Measurement Laboratory, National Institute of Standards and Technology, Gaithersburg, MD 20899, USA

[4]Stanford Synchrotron Radiation Lightsource, SLAC National Accelerator Laboratory, Menlo Park, CA 94025, USA

[5]Cornell High Energy Synchrotron Source, Cornell University, Ithaca, NY 14853, USA

[6]Department of Physics, Pukyong National University, Busan 48513, Republic of Korea

[7]Canadian Institute for Advanced Research, Toronto, Ontario M5G 1Z8, Canada

[*]Corresponding author: takeuchi@umd.edu

## 1. Compositional and optical properties

The $Bi_xNi_{1-x}$ thin-film library shows discontinuous color spectra depending on the Bi composition, indicative of optical phase boundaries. To analyze these boundaries qualitatively and quantitatively, we performed optical characterization using spectroscopic ellipsometry (SE). Figures S1(b) and S1(c) show two-dimensional maps of refractive index ($n$) and extinction coefficient ($k$), respectively, as functions of wavelength and the overall Bi composition on the spread. Both optical coefficient maps show similar contour distributions, where abrupt boundaries are observed at certain Bi compositions, as indicated by black arrows in Fig. S1(b). Figure S1(d) shows the transmittance and reflectance of the $Bi_xNi_{1-x}$ library calculated using the SE spectra. The optical phase boundaries are in good agreement with structural phase boundaries determined by XRD.

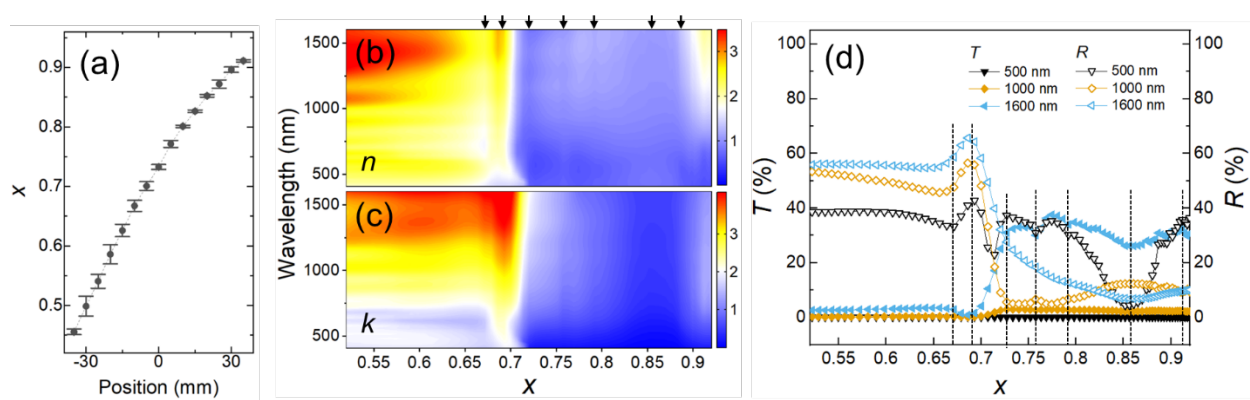

FIG S1. (a) The Bi composition ($x$) mapped across the combinatorial library using WDS, ranging $x \sim 0.5$ to $0.9$. Two-dimensional maps for wavelength-dependent (b) refractive index, $n$, and (c) extinction coefficient, $k$, of the library measured via SE. Abrupt boundaries in the optical constant maps, indicative of optical phase boundaries, are marked with arrows. (d) Transmittance and reflectance of the $Bi_xNi_{1-x}$ combinatorial library calculated from the optical spectra. Dashed lines correspond to the boundaries in (b) and (c).

The full-view and zoom-in-view photograph images of the Bi–Ni thin-film spread are presented in Figs. S2(a) and (b), respectively. The photo images provide clear differences between each phase region, as marked in Fig. 2S(b). The color contrast is particularly significant at $x \approx 0.67$, which corresponds to the phase boundary between Regions A (BiNi) and B (BiNi + Bi$_3$Ni). Figure S3 shows the phase fractions as a function of $x$ obtained from the bulk phase diagram via the lever rule calculation.

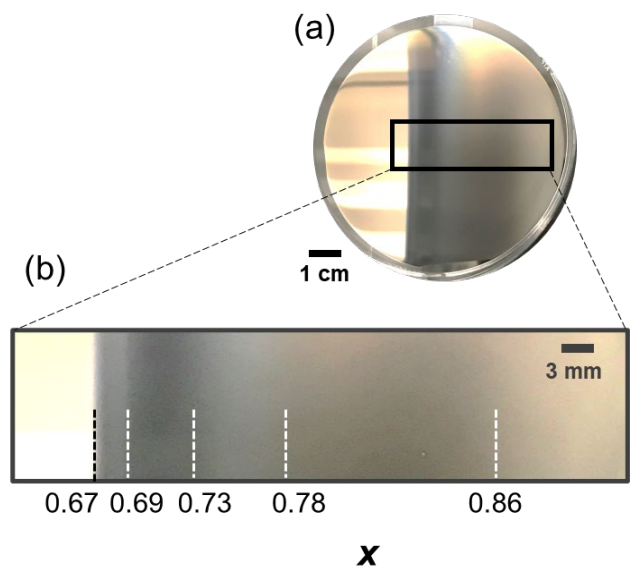

FIG S2. (a) A full-view and (b) zoom-in-view photograph images of the Bi–Ni thin-film spread.

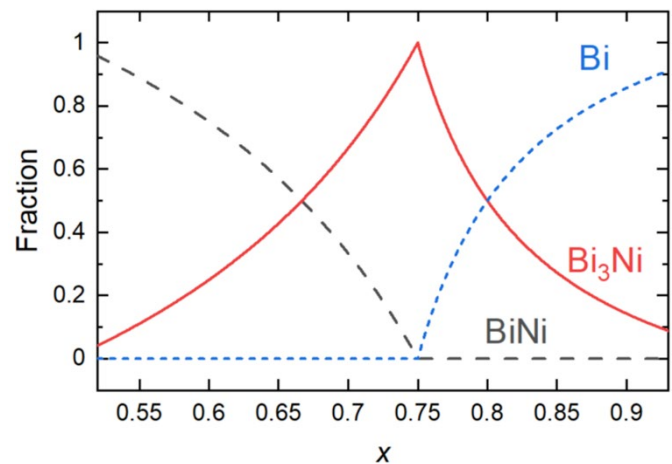

FIG S3. Molar fraction of each phase calculated from the bulk phase diagram using the lever rule.

## 2. Critical temperature as a function of the Bi:Ni ratio

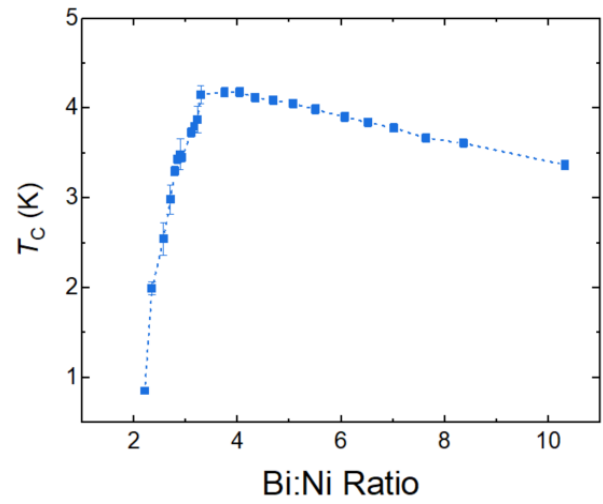

FIG S4. The critical temperature ($T_c$) of $Bi_xNi_{1-x}$ thin films as a function of the Bi:Ni ratio (= $x/(1-x)$).